\documentclass[prx,twocolumn,amsmath,amssymb,superscriptaddress]{revtex4-2}
\usepackage{epsfig, graphicx,graphics,amsmath,amssymb,float}
\usepackage[T1]{fontenc}
\usepackage[latin9]{inputenc}
\usepackage{amsmath}
\usepackage{amssymb}
\usepackage{appendix}
\usepackage{amscd}
\usepackage{bm}
\usepackage{psfrag}
\usepackage{bbm} 
\usepackage{babel}
\usepackage{wasysym }
\usepackage{mathrsfs}
\usepackage{color}
\usepackage[normalem]{ulem}
\usepackage[final]{hyperref} 
\hypersetup{
	colorlinks=true,       
	linkcolor=blue,        
	citecolor=blue,        
	filecolor=magenta,     
	urlcolor=blue         
}
\usepackage{tikz}
\usetikzlibrary{automata, positioning, arrows}
\usepackage{qcircuit}

\definecolor{darkblue}{HTML}{004D6B}
\definecolor{darkred}{HTML}{8c1515}
\definecolor{darkgreen}{HTML}{006400}

\newcommand{\ba}{\begin{array}}
\newcommand{\ea}{\end{array}}
\newcommand{\be}{\begin{equation}}
\newcommand{\ee}{\end{equation}}
\newcommand{\bea}{\begin{eqnarray}}
\newcommand{\eea}{\end{eqnarray}}

\newcommand{\new}[1]{\textcolor{black}{#1}}

\usepackage[american,siunitx]{circuitikz}
\usepackage{amsmath}
\usetikzlibrary{shapes,decorations.pathmorphing}

\begin{document}

\title{Towards scalable active steering protocols for genuinely entangled state manifolds}

\author{Samuel Morales}
\affiliation{Institut f\"ur Theoretische Physik, Heinrich-Heine-Universit\"at, D-40225  D\"usseldorf, Germany}
\author{Silvia Pappalardi}
\affiliation{Institut f\"ur Theoretische Physik, Universit\"at zu K\"oln, Z\"ulpicher Stra{\ss}e 77, 50937 Cologne, Germany}
\author{Reinhold Egger}
\affiliation{Institut f\"ur Theoretische Physik, Heinrich-Heine-Universit\"at, D-40225  D\"usseldorf, Germany}

\begin{abstract} 
We introduce and analyze an active steering protocol designed to target multipartite entangled states. The protocol involves multiple qubits subjected to weak Bell pair measurements with active feedback, where the feedback operations are optimized to maximize the Quantum Fisher Information. Our scheme efficiently reaches a genuinely entangled one-parameter state manifold. Numerical simulations for systems with \new{up to 22 qubits} suggest that the protocol is scalable and allows high multipartite entanglement across the system.
\end{abstract}
\maketitle

\section{Introduction}\label{sec1}

Active steering protocols have recently attracted a lot of attention \cite{Sivak2022,Liu2022a,Friedman2023,Herasymenko2023,Ravindranath2023,Morales2024,Hauser2024,Ackermann2024}. By a sequence of (weak) measurements, followed by feedback operations determined
by the measurement outcomes, one may prepare, stabilize, or manipulate arbitrary quantum states. (Following Ref.~\cite{Roy2020}, we use ``steering'' as proxy for ``guiding'' the system, which differs from ``quantum steering'' in quantum information theory \cite{Uola2020}.)
Weak, i.e., almost non-invasive, measurements can be performed, for instance, by weakly coupling each system qubit to its own detector qubit, with projective measurements of the detector qubits \cite{Wiseman2010}.  
The feedback policy is typically based on a cost function.  
Previous active steering protocols, see, e.g., Refs.~\cite{Herasymenko2023,Morales2024,Ackermann2024}, have been limited to $N\le 6$ system qubits, mainly because practically useful fidelity-based cost functions imply an exponential scaling of the algorithmic demands with system size ($N$).

In this work, we present an active steering protocol that utilizes the \emph{Quantum Fisher 
Information} (QFI) \cite{Helstrom1969} as a cost function. The QFI is a fundamental quantity in entanglement theory and quantum metrology, serving as a witness of multipartite entanglement \cite{giovannetti2011advances, hylluys2012fisher, toth2012multipartite} and a valuable resource for quantum-enhanced metrology \cite{paris2009quantum,pezze2009entanglement,Toth2014,Pezze2018,Liu2020}.
Our QFI-based protocol allows one to efficiently reach a one-parameter manifold of genuinely entangled $N$-qubit states which maximize the QFI, namely Green-Hornberger-Zeilinger (GHZ) states \cite{Nielsen2000}, 
\begin{equation}\label{GHZ}
 |\Psi\rangle=\frac{1}{\sqrt2}\left(|000\cdots\rangle+ e^{i\phi}|111\cdots\rangle\right),
\end{equation}
with an angular parameter $\phi$. (For a system qubit with Pauli 
matrices $\sigma^{x,y,z}$, we use $\sigma^z|0\rangle=|0\rangle$ and $\sigma^z|1\rangle=-|1\rangle$.) By using the QFI as cost function, our protocol significantly accelerates the active steering process. It also allows one to target a specific state with a designated phase $\phi$. Additionally, our findings suggest scalability with the system size $N$, offering promising potential for steering in larger quantum systems.  We note that if one stops the protocol before the maximal value for the QFI has been reached, one may also access more general states beyond Eq.~\eqref{GHZ}.

\new{In contrast to the active steering protocol discussed below, which employs minimally invasive weak measurements, recent works have explored measurement-based protocols with a single round of
strong (projective) Bell measurements plus feedback, see Refs.~\cite{Chen2024nishimori, Sahay2025a,Sahay2025b}
and references therein. Such protocols allow one to prepare a broad family of genuinely multipartite entangled states, including the GHZ state \eqref{GHZ}, in a very efficient manner.  Depending on the experimental platform at hand,  it is nonetheless of interest to study active steering protocols since one can in principle access arbitrary target states, see Ref.~\cite{Morales2024} and our discussion below. Moreover, active steering protocols offer the potential for generalization  to quantum systems with continuous degrees of freedom.}

\new{In Sec.~\ref{sec2},} we formulate the active steering protocol in a platform-independent manner.  
Weak measurements and the associated quantum feedback effects have by now become standard experimental tools which are employed in various platforms, see, e.g., 
Refs.~\cite{Hosten2008,Dixon2009,Palacios2010,Riste2013,Groen2013,Zhang2017,Minev2019,Cujia2019,Kim2021a}. They allow for high-precision measurements, and an application of these techniques to our protocol could be very promising.  Moreover, as shown recently in Ref.~\cite{Ackermann2024} \new{for one or two system qubits,} by including amplitude damping and dephasing in the stochastic master equation, active steering schemes of the type considered below are \new{expected to be} robust against the presence of error channels with sufficiently weak error rate. In fact, Ref.~\cite{Ackermann2024} established the existence of a finite threshold  error rate, and as long as the error rate stays below the threshold, errors can be corrected ``on the fly'' by the protocol.  \new{Albeit numerical simulations of the present version of our protocol including error channels are computationally prohibitively expensive for large $N$ because one has to simulate the time evolution of mixed states, the fact that the error threshold rate is almost identical for $N=1$ and $N=2$ \cite{Ackermann2024} suggests that a finite error threshold will persist at least for moderate values of $N$. 
If the system dynamics is described by a mixed state, one has to implement 
an unraveling procedure in practice for numerical stability.  For large $N$, such approaches become
exponentially costly in terms of computational demands.
This restriction also applies when the detectors have non-ideal measurement efficiency 
since such effects are also captured by describing the system dynamics in terms of mixed states
\cite{Jacobs2006}.   Furthermore, the QFI becomes more complicated for mixed states.
For simplicity, we thus neglect external noise channels and non-ideal measurement efficiencies, and 
study the error-free case with ideal measurements below. However, a modified implementation of our
protocol that may allow to circumvent these restrictions is discussed in Sec.~\ref{sec4}. }

Since weak measurements play a key role in our protocol, a physical realization with fast qubit readout is desirable, e.g., superconducting Andreev qubits with detector readout times of order 10 ns \cite{Janvier2015} could offer a good option \cite{Pita2025}.
We note that active steering protocols of similar type have recently been experimentally implemented  \cite{Volya2023,Volya2024}. 
Our protocol assumes that the system is initialized at time $t=0$ 
in a simple product state, say, ${|\Psi(t=0)\rangle=|000\cdots\rangle}$, and that for a given measurement record, 
the state trajectory $|\Psi(t)\rangle$ can be stored and updated on a classical computer for each time step of the quantum protocol. 
\new{After introducing the protocol in Sec.~\ref{sec2}, we present numerical simulation results in Sec.~\ref{sec3}. We  primarily focus on the GHZ state, but we also show how to prepare so-called Dicke states using this approach.    
In Sec.~\ref{sec4}, we discuss several open issues and directions for future research.
In particular, we outline how the state tracking requirement may be avoided in modified implementations of our protocol.  }

\section{Protocol and QFI}\label{sec2}
 
\begin{figure}[t]
\centering 
\footnotesize
            \begin{tikzpicture}[scale=2]
            \node[] at (-0.25,1.5) (a) {\large (a)};
            \node[] at (2,-1.55) (b) {\large space};
                \node[state, draw=gray!50, fill=gray!50, minimum size=3mm] at (0.5,1.2) (d11) {};
                \node[rectangle, draw=red!60, fill=red!60, minimum size=3mm] at (0.5,0.8) (s11) {};
                \draw [-, black] (d11) edge[thick] node[left]{} (s11);
                
                \node[state, draw=gray!50, fill=gray!50, minimum size=3mm] at (1,1.2) (d21) {};
                \node[rectangle, draw=red!60, fill=red!60, minimum size=3mm] at (1,0.8) (s21) {};
                \draw [-, black] (d21) edge[thick] node[left]{} (s21);
                
                \node[state, draw=gray!50, fill=gray!50, minimum size=3mm] at (1.5,1.2) (d31) {};
                \node[rectangle, draw=red!60, fill=red!60, minimum size=3mm] at (1.5,0.8) (s31) {};
                \draw [-, black] (d31) edge[thick] node[left]{} (s31);
                
                \node[state, draw=gray!50, fill=gray!50, minimum size=3mm] at (2,1.2) (d41) {};
                \node[rectangle, draw=red!60, fill=red!60, minimum size=3mm] at (2,0.8) (s41) {};
                \draw [-, black] (d41) edge[thick] node[left]{} (s41);

                \node[state, draw=gray!50, fill=gray!50, minimum size=3mm] at (2.5,1.2) (d51) {};
                \node[rectangle, draw=red!60, fill=red!60, minimum size=3mm] at (2.5,0.8) (s51) {};
                \draw [-, black] (d51) edge[thick] node[left]{} (s51);

                \node[state, draw=gray!50, fill=gray!50, minimum size=3mm] at (3,1.2) (d61) {};
                \node[rectangle, draw=red!60, fill=red!60, minimum size=3mm] at (3,0.8) (s61) {};
                \draw [-, black] (d61) edge[thick] node[left]{} (s61);
                
                \node[] at (0.75,1.5) {\includegraphics[scale=.25]{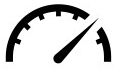}};
                \draw [thick, fill=gray!30] (0.5,1.325) to [out=150,in=30] (1,1.325) to [out=160,in=20] (0.5,1.325);
                \node[] at (1.75,1.5) {\includegraphics[scale=.25]{dial.png}};
                \draw [thick, fill=gray!30] (1.5,1.325) to [out=150,in=30] (2,1.325) to [out=160,in=20] (1.5,1.325);
                \node[] at (2.75,1.5) {\includegraphics[scale=.25]{dial.png}}; 
                \draw [thick, fill=gray!30] (2.5,1.325) to [out=150,in=30] (3,1.325) to [out=160,in=20] (2.5,1.325);
                
                \node[state, draw=gray!50, fill=gray!50, minimum size=3mm] at (0.5,0.2) (d12) {};
                \node[rectangle, draw=red!60, fill=red!60, minimum size=3mm] at (0.5,-0.2) (s12) {};
                \draw [-, black] (d12) edge[thick] node[left]{} (s12);
                
                \node[state, draw=gray!50, fill=gray!50, minimum size=3mm] at (1,0.2) (d22) {};
                \node[rectangle, draw=red!60, fill=red!60, minimum size=3mm] at (1,-0.2) (s22) {};
                \draw [-, black] (d22) edge[thick] node[left]{} (s22);
                
                \node[state, draw=gray!50, fill=gray!50, minimum size=3mm] at (1.5,0.2) (d32) {};
                \node[rectangle, draw=red!60, fill=red!60, minimum size=3mm] at (1.5,-0.2) (s32) {};
                \draw [-, black] (d32) edge[thick] node[left]{} (s32);
                
                \node[state, draw=gray!50, fill=gray!50, minimum size=3mm] at (2,0.2) (d42) {};
                \node[rectangle, draw=red!60, fill=red!60, minimum size=3mm] at (2,-0.2) (s42) {};
                \draw [-, black] (d42) edge[thick] node[left]{} (s42);

                \node[state, draw=gray!50, fill=gray!50, minimum size=3mm] at (2.5,0.2) (d52) {};
                \node[rectangle, draw=red!60, fill=red!60, minimum size=3mm] at (2.5,-0.2) (s52) {};
                \draw [-, black] (d52) edge[thick] node[left]{} (s52);

                \node[state, draw=gray!50, fill=gray!50, minimum size=3mm] at (3,0.2) (d62) {};
                \node[rectangle, draw=red!60, fill=red!60, minimum size=3mm] at (3,-0.2) (s62) {};
                \draw [-, black] (d62) edge[thick] node[left]{} (s62);
                
                \node[] at (0.25,0.5) {\includegraphics[scale=.25]{dial.png}};
                \draw [thick, fill=gray!30] (0,0.325) to [out=150,in=30] (0.5,0.325) to [out=160,in=20] (0,0.325);
                \node[rectangle, draw=white, fill=white, minimum size=6mm] at (0.1,0.46) (dial1) {};
                \node[] at (1.25,0.5) {\includegraphics[scale=.25]{dial.png}};
                \draw [thick, fill=gray!30] (1,0.325) to [out=150,in=30] (1.5,0.325) to [out=160,in=20] (1,0.325);
                \node[] at (2.25,0.5) {\includegraphics[scale=.25]{dial.png}};
                \draw [thick, fill=gray!30] (2,0.325) to [out=150,in=30] (2.5,0.325) to [out=160,in=20] (2,0.325);
                \node[] at (3.25,0.5) {\includegraphics[scale=.25]{dial.png}};
                \draw [thick, fill=gray!30] (3,0.325) to [out=150,in=30] (3.5,0.325) to [out=160,in=20] (3,0.325);
                \node[rectangle, draw=white, fill=white, minimum size=6mm] at (3.4,0.46) (dial2) {};

                \node[state, draw=gray!50, fill=gray!50, minimum size=3mm] at (0.5,-0.8) (d13) {};
                \node[rectangle, draw=red!60, fill=red!60, minimum size=3mm] at (0.5,-1.2) (s13) {};
                \draw [-, black] (d13) edge[thick] node[left]{} (s13);
                
                \node[state, draw=gray!50, fill=gray!50, minimum size=3mm] at (1,-0.8) (d23) {};
                \node[rectangle, draw=red!60, fill=red!60, minimum size=3mm] at (1,-1.2) (s23) {};
                \draw [-, black] (d23) edge[thick] node[left]{} (s23);
                
                \node[state, draw=gray!50, fill=gray!50, minimum size=3mm] at (1.5,-0.8) (d33) {};
                \node[rectangle, draw=red!60, fill=red!60, minimum size=3mm] at (1.5,-1.2) (s33) {};
                \draw [-, black] (d33) edge[thick] node[left]{} (s33);
                
                \node[state, draw=gray!50, fill=gray!50, minimum size=3mm] at (2,-0.8) (d43) {};
                \node[rectangle, draw=red!60, fill=red!60, minimum size=3mm] at (2,-1.2) (s43) {};
                \draw [-, black] (d43) edge[thick] node[left]{} (s43);

                \node[state, draw=gray!50, fill=gray!50, minimum size=3mm] at (2.5,-0.8) (d53) {};
                \node[rectangle, draw=red!60, fill=red!60, minimum size=3mm] at (2.5,-1.2) (s53) {};
                \draw [-, black] (d53) edge[thick] node[left]{} (s53);
                
                \node[state, draw=gray!50, fill=gray!50, minimum size=3mm] at (3,-0.8) (d63) {};
                \node[rectangle, draw=red!60, fill=red!60, minimum size=3mm] at (3,-1.2) (s63) {};
                \draw [-, black] (d63) edge[thick] node[left]{} (s63);
                
                \node[] at (0.75,-0.5) {\includegraphics[scale=.25]{dial.png}};
                \draw [thick, fill=gray!30] (0.5,-0.675) to [out=150,in=30] (1,-0.675) to [out=160,in=20] (0.5,-0.675);
                \node[] at (1.75,-0.5) {\includegraphics[scale=.25]{dial.png}};
                \draw [thick, fill=gray!30] (1.5,-0.675) to [out=150,in=30] (2,-0.675) to [out=160,in=20] (1.5,-0.675);
                \node[] at (2.75,-0.5) {\includegraphics[scale=.25]{dial.png}};
                \draw [thick, fill=gray!30] (2.5,-0.675) to [out=150,in=30] (3,-0.675) to [out=160,in=20] (2.5,-0.675);
        
                \draw [->] (0.25,-1.35) -- node[left]{{\LARGE $t$}} (0.25,1.5);
                \draw [->] (0.3,-1.4) -- node{} (3.3,-1.4);
            \end{tikzpicture}
        \begin{tikzpicture}[scale=1.8]
    \node[] at (-1,1) (a) {\large (b)};

    \foreach \i in {0,...,5}
    {
        \pgfmathtruncatemacro{\label}{\i}
        \node[state, draw=gray!30, fill=gray!30, minimum size=3mm] (\label) at ({sin(deg(\i*2*pi/6))}, {cos(deg(\i*2*pi/6))}) {};
    }
    \foreach \i in {0,...,4}
    {
    \pgfmathtruncatemacro{\next}{\i + 1}
    \draw (\i)--({\next}) ;
    }
    \draw (5)--(0);
\end{tikzpicture}
\begin{tikzpicture}[scale=1.8]
    \node[] at (-1,1) (a) {\large (c)};
    \foreach \i in {0,...,5}
    {
        \pgfmathtruncatemacro{\label}{\i}
        \node[state, draw=gray!30, fill=gray!30, minimum size=3mm] (\label) at ({sin(deg(\i*2*pi/6))}, {cos(deg(\i*2*pi/6))}) {};
    }
    \foreach \i in {0,...,5}
    \foreach \j in {0,...,5}
    {
    \draw (\i)--(\j) ;
    }
\end{tikzpicture}
\caption{\new{Schematic time evolution of the active steering protocol. (a) We show three time steps for $N=6$ system qubits (red squares) coupled by steering operators $H_n$ (straight vertical lines) to their own detector qubits (grey circles).  The qubit chain has periodic boundary conditions.  A possible scheme for the Bell measurements of neighboring detector qubit pairs in subsequent cycles is indicated.  (b) We mainly consider the case of Bell measurements of nearest-neighbor detector qubit pairs. (c) In Fig.~\ref{fig5} below, we also study the case where arbitrary detector qubit pairs can be subjected to Bell measurements (full connectivity).} }
\label{fig1}
\end{figure}

\begin{table}[t]
    \centering
        choose $|\Psi(0)\rangle=|0\dots0\rangle$ and $\mathcal{O}$\\\smallskip
    \begin{ruledtabular}
        \begin{tabular}{ccc}
             \underline{Classical computer} & &  \underline{Quantum computer}\\
             &&\\
             $|\Psi(t)\rangle$ & & $|\Psi(t)\rangle\rightarrow |\Psi(t)\rangle\otimes |00\rangle_d$\\
             $\downarrow$ & & $\downarrow$\\
             $\mathrm{max}_{K\in\mathcal{K}} \langle dF_Q(|\Psi\rangle,K)\rangle_{\rm ms}$ & $\overset{K\in\mathcal{K}}{\longrightarrow}$ & $e^{-i\delta t H_K}|\Psi(t)\rangle\otimes |00\rangle_d$ \\
             $\downarrow$ & & $\downarrow$\\
             $|\Psi(t+\delta t|\xi,\eta)\rangle$ & $\xleftarrow[]{(\xi,\eta)}$ & ${}_d\langle\Phi_{\xi,\eta}|e^{-i\delta t H_K}|00\rangle_d \times |\Psi(t)\rangle$
        \end{tabular}
  \end{ruledtabular}      
    \caption{\new{Active steering protocol using the QFI as cost function. A classical computation of the measurement-averaged cost function change $\langle dF_Q(|\Psi\rangle,K)\rangle_{\rm ms}$ determines the optimal coupling $K=(K_n, K_{n+1})$ out of the coupling family  $\mathcal{K}=\{K_n=(\alpha_n,\beta_n)|\alpha_n\in\{x,y,z\}; \beta_n\in\{x,z\}\}$ for qubit pair $(n, n+1)$. This choice is fed into the quantum computer as system-detector coupling. After unitary time evolution of duration $\delta t$, the detector is measured in the Bell basis $|\Phi_{\xi,\eta}\rangle_d$, where $\xi$ and $\eta$ correspond to the possible measurement outcomes. These are fed back into the classical computer to update and keep track of the system state.}}
    \label{tab1}
\end{table}

We schematically illustrate the protocol in Fig.~\ref{fig1} and Table \ref{tab1}.
\new{Each protocol step of time duration $\delta t$ has two
components, namely (i) unitary evolution of the coupled system and detector qubits under the chosen
feedback Hamiltonian, followed by (ii) weak measurements of the system qubits via Bell pair measurements of the detector
qubits.
Depending on the measurement outcomes, the feedback Hamiltonian for the next iteration step is then determined according 
to the decision making scheme
discussed below.}

We consider $N$  system qubits described by Pauli matrices $\sigma_n^\alpha$ (with $\alpha=x,y,z$ and $n=1,\ldots,N$), where each system qubit couples only to its own detector qubit described by Pauli matrices $\tau_n^\beta$.  We neither allow for direct couplings between different system qubits nor between different detector qubits, while the Hamiltonian $H_n$ (``steering operator'') for the $n$th system-detector qubit pair can be selected from the set of Pauli gates,
\begin{equation}
    \label{steeringop}
    H_n = J \sigma_n^{\alpha_n} \tau_n^{\beta_n} ,
\end{equation}
with $\alpha_n \in \{x,y,z\}$ and $\beta_n\in \{x,z\}$.  For simplicity,
we assume a fixed coupling $J$, \new{where we put $J=+1$ in what follows}, 
and degenerate zero-energy states for all uncoupled qubits.
The steering parameters $K_n=(\alpha_n,\beta_n)$ are chosen according to a decision making scheme in every time step of the protocol as described below, with $[H_n,H_{n'}]=0$ for arbitrary $K_n$ and $K_{n'}$.

\new{Table \ref{tab1} illustrates the active steering protocol using Bell state measurements on  neighboring pairs, see Fig.~\ref{fig1}(b), of detector qubits.  Later on, we also consider the case
in Fig.~\ref{fig1}(c) where Bell measurements can be performed for arbitrarily chosen pair orderings (where a given pairing order is determined from a uniform random distribution), but we focus on the nearest-neighbor case in what follows. For a given time step of the protocol,  all non-overlapping pairs $(n,n+1)$ can be steered simultaneously, see Fig.~\ref{fig1}(a).  One can either choose an alternating sequence of pairings on subsequent time steps, as shown in Fig.~\ref{fig1}(a), or simply assign the pairing order randomly.  In our simulations, we found the second option to be more efficient.  As explained below,
the active decision making is performed on a classical computer using the knowledge about the present state of the system.}

\new{Let us now describe the protocol in detail. We start}
at time $t=0$ by initializing all system and detector qubits in $|0\rangle$ and $|0\rangle_d$ (the subscript $d$ refers to detector qubit space), respectively, i.e., the system state is $|\Psi(t=0)\rangle=|000\cdots\rangle$.
We then group the $N$ qubits into neighboring pairs ($n,n+1$), see Fig.~\ref{fig1}, where all subsequent operations for different pairs commute and can thus be performed simultaneously. (For odd $N$, one ``idle'' qubit remains whose location is chosen from a uniform random distribution.)
Given the state $|\Psi(t)\rangle$, \new{for active steering towards the GHZ state}, we select the steering couplings for this pair, $(K_n,K_{n+1})$, such that the measurement-averaged expectation value of the QFI after a time step of duration $\delta t$ is maximized. 

For a pure $N$-qubit state $|\Psi\rangle$, the QFI is defined as \cite{paris2009quantum,pezze2009entanglement,Toth2014,Pezze2018,Liu2020} 
\begin{equation}\label{QFI}
 F_Q = 4\left(\langle \Psi| {\cal O}^2|\Psi \rangle - \langle \Psi|{\cal O}|\Psi\rangle^2 \right) .
\end{equation}
For collective observables ${\cal O}=\frac12
 \sum_{n=1}^N O_n$, where $O_n$ are local operators, the QFI can be used to probe the multipartite entanglement structure of the state $|\Psi\rangle$  \cite{giovannetti2011advances, hylluys2012fisher, toth2012multipartite}. If the QFI satisfies the inequality $F_Q >m N$, then at least $(m+1)$ parties of the system are entangled. Namely, $m\leq N$ represents the size of the biggest entangled block. The upper bound $F_Q\sim N^2$ corresponds to the so-called genuinely multipartite entanglement.
\new{In particular, the family of states which saturate the
maximum value of the multipartite entanglement is an arbitrary superposition of the
eigenvectors of ${\cal O}$ with largest and smallest
eigenvalues. For ${\cal O}=\frac12 \sum_{n=1}^N \sigma^z_n$, these correspond to the states defined
in Eq.~\eqref{GHZ}.}

 In our protocol, we parametrize $O_n={\bf s}_n\cdot \boldsymbol{\sigma}_n$, with ${{\bf s}_n=(s_n^x,s_n^y,s_n^z)}$ an arbitrary unit vector and ${\boldsymbol{\sigma}_n=(\sigma_n^x,\sigma_n^y,\sigma_n^z)}$. 
Using the optimal choice for $(K_n,K_{n+1})$, \new{discussed below after Eq.~\eqref{dfq},}
one time-evolves the coupled system-plus-detector system for a time step $\delta t$. 
Next, a projective measurement of the detector qubit pair is done in its Bell basis $\{|\Phi_{\xi,\eta}\rangle_d\}$ \cite{Nielsen2000}, where $|\Phi_{\xi=0,\eta=\pm}\rangle_d = \left(|00\rangle_d \pm |11\rangle_d\right) /\sqrt2$ and
$|\Phi_{\xi=1,\eta=\pm}\rangle_d = \left(|01\rangle_d \pm |10\rangle_d\right)/\sqrt2$. Symmetric ($\eta=+1$) and antisymmetric ($\eta=-1$) Bell states have even $(\xi=0)$ or odd ($\xi=1$) parity, where ``symmetry'' refers to qubit exchange while even (odd) ``parity'' means that states are built from the basis $\{ |00\rangle_d, |11\rangle_d\}$  ($\{|01\rangle_d, |10\rangle_d$\}). 
Such measurements can be implemented by commuting measurements of the Pauli operators $\tau_n^z \tau_{n+1}^{z}=\pm 1$ and $\tau_n^x \tau_{n+1}^{x}=\pm 1$. Finally, one re-initializes all detector qubits in the state $|0\rangle_d$ and iterates the protocol until convergence has been achieved. Since the initial detector state $|00\rangle_d$ (for each time step and each qubit pair) has even parity, measurement outcomes with odd parity ($\xi=1$) are 
referred to as quantum jumps. The above measurements realize entanglement swapping \cite{Boschi1998,Pan1998,Jennewein2001,Gisin2005,Riebe2008,Kaltenbaek2009,Horodecki2009,Huang2023} and 
tend to increase entanglement in the system state $|\Psi(t)\rangle\to |\Psi(t+\delta t)\rangle$, see Ref.~\cite{Morales2024} for a detailed discussion.

\begin{figure}
    \centering
    \includegraphics[width=\linewidth]{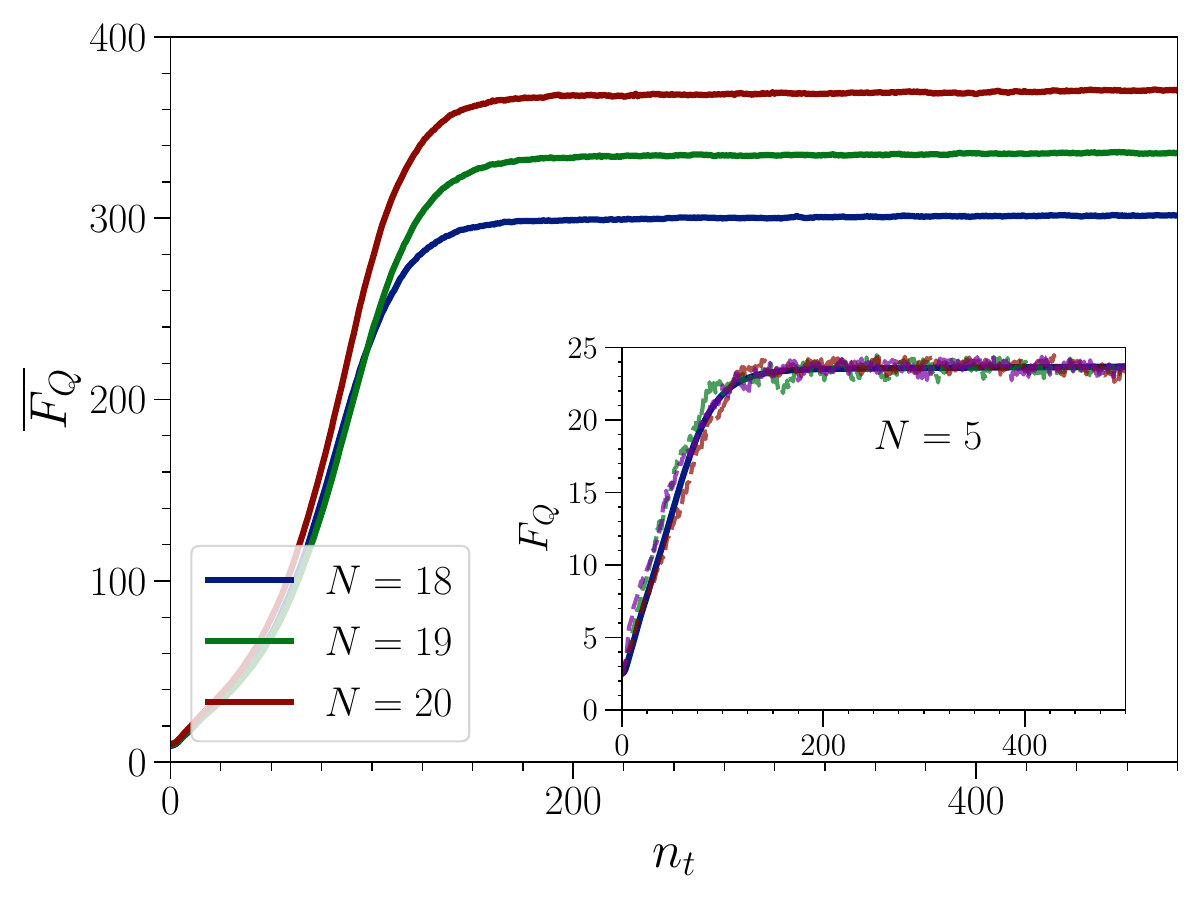}
    \caption{Average QFI $\overline{F_Q}$ vs number of time steps $n_t=t/\delta t$ for different $N$ and $J\delta t=0.2$. Note that $F_Q\le N^2$. Averages are over $100$ trajectories. The inset shows $\overline{F_Q}$ vs $n_t$ (solid curve) for $N=5$ (averaged over 8000 trajectories), together with three individual measurement trajectories (dashed curves).
   }    
    \label{fig2}
\end{figure}

\begin{figure*}
    \centering
    \includegraphics[width=0.3\linewidth]{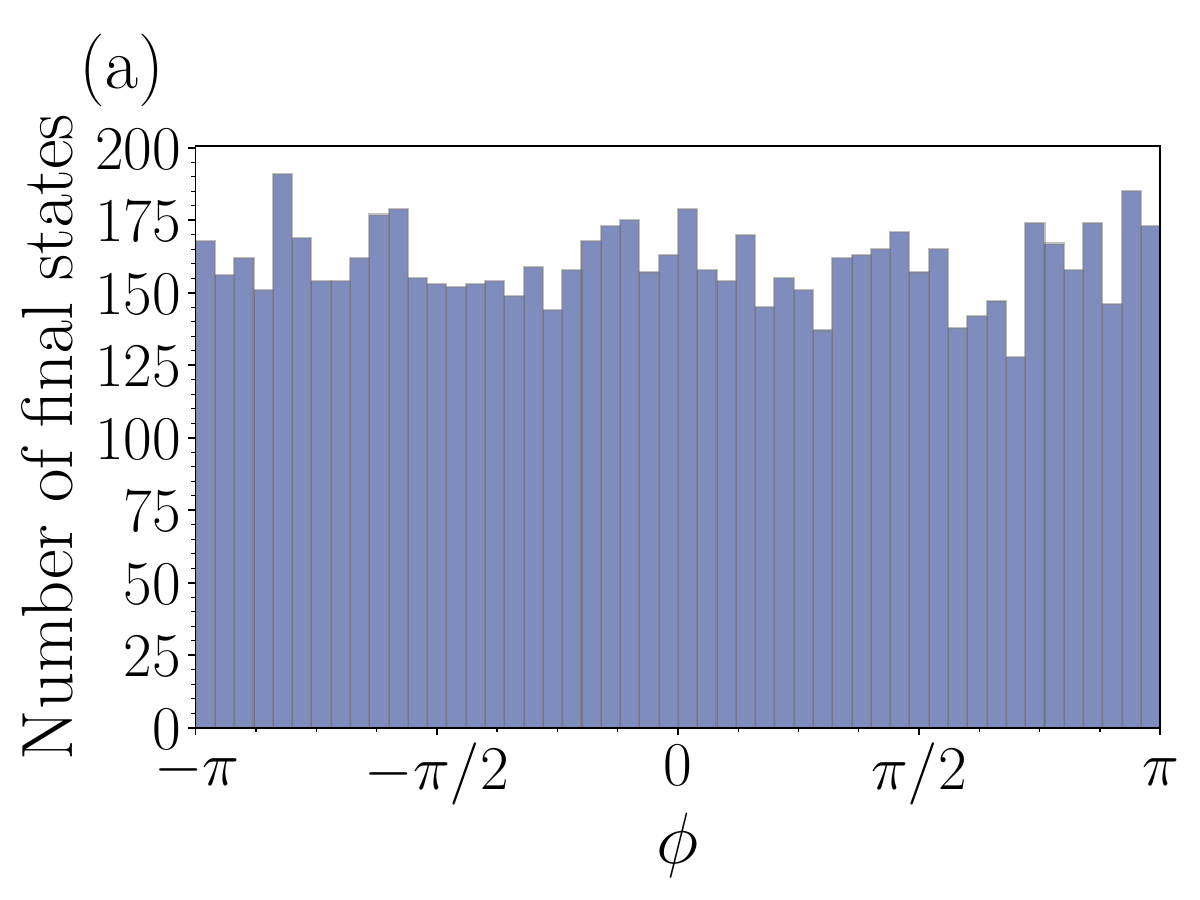}
    \includegraphics[width=0.3\linewidth]{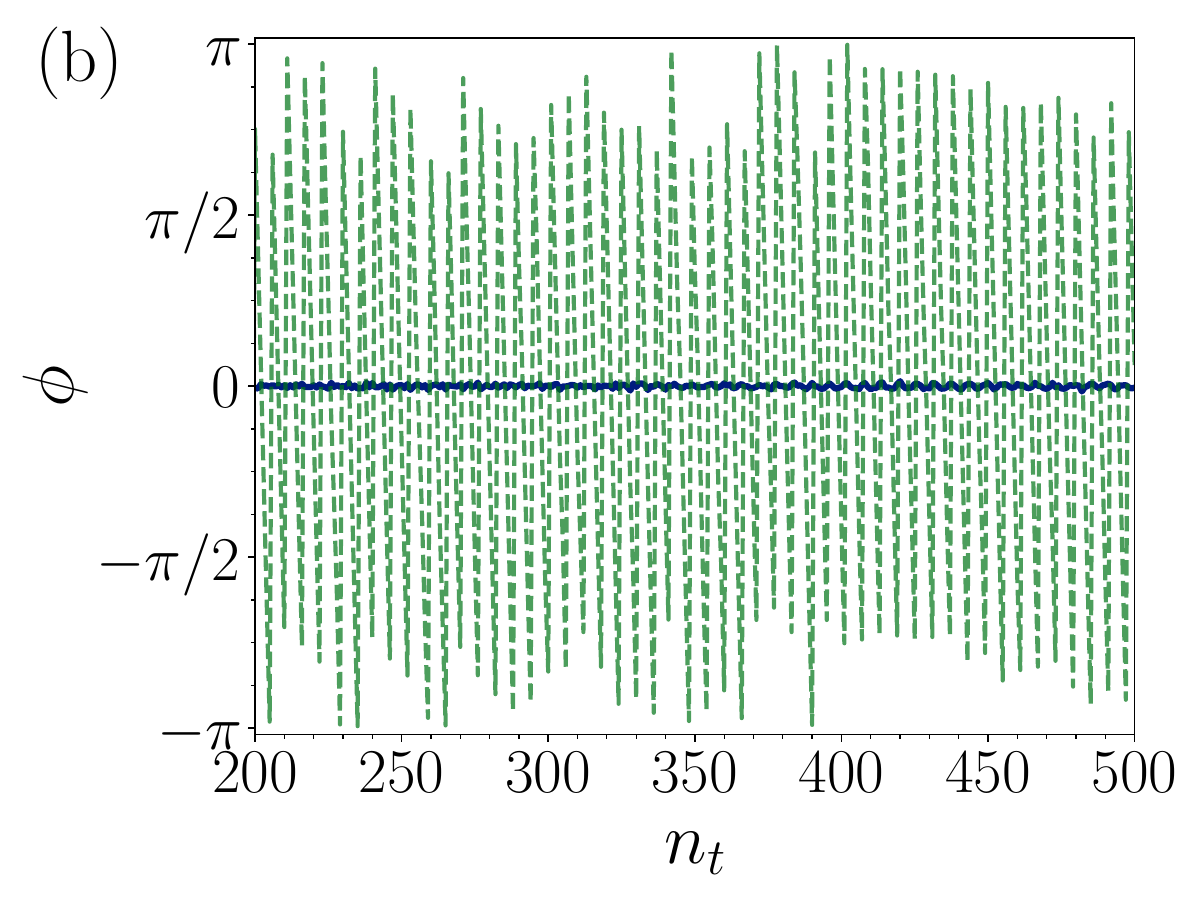}
    \includegraphics[width=0.3\linewidth]{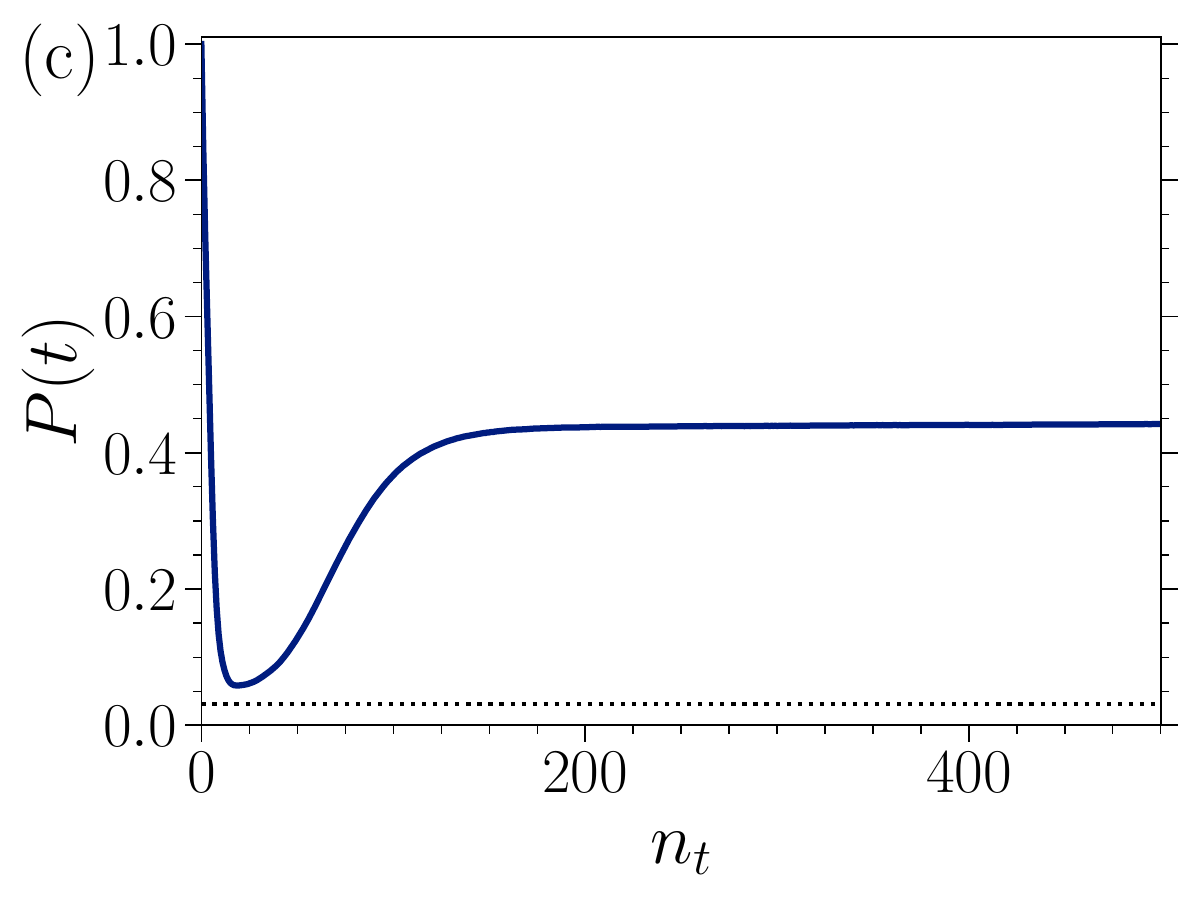}
    \caption{Active steering protocol for $N=5$ and $J\delta t=0.2$. Measurement averages are over $8000$ trajectories.  (a) Histogram of the phase $\phi$ in Eq.~\eqref{GHZ} found after $n_t=500$ time steps.  (b) Time evolution of the phase $\phi$ after convergence to the target manifold has been reached, shown for one individual measurement trajectory (dashed green curve) and for the average  (solid blue curve).  (c)  Purity $P(t)$ vs  $n_t$, see Eq.~\eqref{purity}. The dotted line corresponds to the maximally mixed (infinite temperature) state. }
    \label{fig3}
\end{figure*}

In the weak measurement limit $J\delta t \ll 1$ \cite{Wiseman2010}, the state change $|d\Psi\rangle=|\Psi(t+\delta t)\rangle-|\Psi\rangle$ with $|\Psi\rangle=|\Psi(t)\rangle$ for measurement outcome $(\xi,\eta)$ is governed by a jump-type nonlinear stochastic Schr\"odinger equation (SSE)
\cite{Breuer2002,Jacobs2006,Zhang2017},
\begin{eqnarray} \nonumber
|d\Psi\rangle &=&\Biggl [ -i\delta t H_0+\xi \Biggl( \frac{c_{\eta}^{}}{\sqrt{\langle c_{\eta}^\dagger c_{\eta}^{} \rangle}}-1\Biggr)    \label{SSE}
 \\&& \qquad 
- \frac{\delta t}{2}  \left(c^\dagger_{\eta} c^{}_{\eta} - \langle c_{\eta}^\dagger c_{\eta}^{} \rangle\right) 
\Biggr] |\Psi\rangle,
\end{eqnarray}
where $H_0 = J \sum_{m=n,n+1} \delta_{\beta_m,z} \sigma_m^{\alpha_m}$ and  $\langle c^\dagger_\eta c^{}_\eta\rangle = \langle \Psi|c^\dagger_\eta c^{}_\eta|\Psi\rangle$.
The jump operators $c_{\eta=\pm}$ are given by 
\begin{equation}\label{jumpops}
   c_\eta = -iJ\,\sqrt{\delta t}\left( \eta \delta_{\beta_n,x} \sigma_n^{\alpha_n}  +  
   \delta_{\beta_{n+1},x} \sigma_{n+1}^{\alpha_{n+1}}\right),
\end{equation}
where the outcome $(\xi,\eta)$ has the \emph{a priori} probability
$p^{}_{\xi,\eta}=\frac12[\delta_{\xi,0} + (\delta_{\xi,1}-\delta_{\xi,0}) 
\delta t \langle c^\dagger_\eta c^{}_\eta\rangle]$.  Averaging over the measurement outcomes
after one time step, one arrives at the anticipated average change in QFI after the next time step,
\begin{eqnarray}\label{dfq}
    \langle dF_Q \rangle_{\rm ms} &=&\langle F_Q(t+\delta t)\rangle_{\rm ms}-F_Q(t) \\ \nonumber
    &=& 4\Bigl( \mathrm{Tr}\left({\cal O}^2\, \langle d\rho \rangle_{\rm ms}\right)-2\mathrm{Tr}\left( {\cal O}\, \langle d\rho\rangle_{\rm ms}
    \right)\mathrm{Tr}({\cal O} \rho) \\ 
    &-&\left\langle\left[\mathrm{Tr}({\cal O} \,d\rho)\right]^2\right\rangle_{\rm ms} \Bigr)\nonumber
\end{eqnarray} 
with $d\rho=|d\Psi\rangle\langle \Psi|+|\Psi\rangle\langle d\Psi|+|d\Psi\rangle\langle d\Psi|$ and ${\rho=|\Psi\rangle\langle \Psi|}$. (Here $\langle A\rangle_{\rm ms}$ denotes a measurement average of the quantity $A$ using the probabilities $p^{}_{\xi,\eta}$.)   
We compute $\langle dF_Q  \rangle_{\rm ms}$ for all possible steering parameters and then choose 
$(K_n,K_{n+1})$ such that $\langle dF_Q \rangle_{\rm ms}$ is maximized.  
\new{We emphasize that $(K_n,K_{n+1})$ is determined separately for each qubit pair at a given time step.}
In our numerical simulations, the system state is propagated according to the SSE \eqref{SSE}.  In contrast to Ref.~\cite{Morales2024}, the time evolution is not terminated once a certain threshold value for $F_Q$ has been reached but the quantum state trajectory continues evolving according to the above protocol.

\new{Let us provide some details on the calculation of $\langle dF_Q\rangle_{\rm ms}$.
The classical computation of $\langle dF_Q(|\Psi\rangle,K)\rangle_{\rm ms}$ for $H_K=H_n+H_{n+1}$ with $K=(K_n,K_{n+1})$ is performed numerically in a  Bloch tensor representation of the system state,
\begin{equation}\label{Bloch}
  |\Psi\rangle\langle\Psi| = \frac{1}{2^N} \sum_{\cal S} R_{\cal S} {\cal S}, 
  \quad R_{\cal S}=\langle\Psi|{\cal S}|\Psi\rangle,
\end{equation}
with the Pauli string operator ${\cal S} = \sigma_1^{\mu_1} \sigma_2^{\mu_2} \cdots \sigma_N^{\mu_N}$ with $\mu_j\in\{0,1,2,3\}$.
Using this representation to parametrize the observables, the average QFI change is obtained as
\begin{widetext}
\begin{align}\label{eqApp:expQFIchg}
    \langle dF_Q\rangle_{\rm ms}&=\sum_{n\neq m}\sum_{\alpha_i,\alpha_j} s_n^{\alpha_n}s_m^{\alpha_m} \langle dQ_{n,m}^{\alpha_n,\alpha_m}\rangle_{\rm ms}
    -2\left(\sum_{n,\alpha_n} s_n^{\alpha_n} \langle dR_n^{\alpha_n}\rangle_{\rm ms}\right)\left(\sum_{n,\alpha_n} s_n^{\alpha_n} R_n^{\alpha_n}\right) 
    -2\delta t\sum_\eta \frac{1}{\langle c_\eta^\dagger c_\eta \rangle}\left(\sum_{n,\alpha_n} s_n^{\alpha_n} G^{(\eta)}_{\alpha_n}\right)^2,
\end{align}
\end{widetext}
where we used the reduced single-qubit Bloch vectors $R_n^{\alpha_n}=\langle\Psi|\sigma_n^{\alpha_n}|\Psi\rangle$ and the two-qubit correlators $Q_{n,m}^{\alpha_n,\alpha_m}=\langle\Psi|\sigma_n^{\alpha_n}\sigma_m^{\alpha_m}|\Psi\rangle$. For explicit expressions for $\langle dR_n^{\alpha_n}\rangle_{\rm ms}$ and $\langle dQ_{n,m}^{\alpha_n,\alpha_m}\rangle_{\rm ms}$, see Eq.~(27) in Ref.~\cite{Morales2024}. The second-order state change of the single qubit density matrices for $\xi=1$ measurement outcomes is given by
\begin{widetext}
\begin{align}
    G^{(\eta)}_{\mu_i}&= -\sum_{m=n,n+1} \sum_{\alpha\ne \alpha_i} \Gamma_m \delta_{\beta_m,x} \delta_{\mu_i,\alpha}  R_{\mu_i} + \eta \sqrt{\Gamma_n\Gamma_{n+1}}  \delta_{\beta_n,x}\delta_{\beta_{n+1},x} \left(\mathcal{F}_{\mu_i}-Q_{n,n+1} R_{\mu_i}\right), \\ \nonumber
    \mathcal{F}_{\cal S} &= \frac{1}{2^{N+1}} \sum_{{\cal S}'} R_{{\cal S}'} {\rm Tr}\left( 
    (\sigma_n^{\alpha_n} {\cal S}' \sigma_{n+1}^{\alpha_{n+1}} + \sigma_{n+1}^{\alpha_{n+1}} {\cal S}' \sigma_n^{\alpha_n} ) {\cal S}\right).
\end{align}
\end{widetext}
With a chosen coupling $(K_n,K_{n+1})$ for the following time step, the quantum system is
evolved accordingly, and the detector is measured subsequently. 
The stochastic outcome of this measurement is recorded and used to keep track of the running state on the classical computer.}

\new{We note that for the preparation of non-GHZ state families, one may choose a specific target value of the QFI, $F_Q^*$, and then 
minimize the cost function $|F_Q-F_Q^\ast|$. This approach allows one to use active steering protocols targeting different classes of state manifolds associated with a specific QFI value.  
As an example, we show results for Dicke states in Sec.~\ref{sec3} below.}

\section{Numerical simulation results}\label{sec3}

We now show simulation results for the above protocol maximizing the QFI. For the steering operator set \eqref{steeringop}, it is convenient to choose ${\bf s}_n=\frac{1}{\sqrt2}(1,0,1)^T$, see Eq.~\eqref{QFI}, but our results are robust under small rotations of this unit vector.  Correspondingly, the GHZ states \eqref{GHZ} are defined with respect to rotated states $|0\rangle\to |0'\rangle$ and $|1\rangle\to |1'\rangle$. Up 
to a normalization factor, $|0'\rangle=|0\rangle+(\sqrt2-1)|1\rangle$ and $|1'\rangle=(1-\sqrt2)|0\rangle + |1\rangle$.   
In Fig.~\ref{fig2}, we show the evolution of the time-dependent QFI. The main panel illustrates the average QFI $\overline{F_Q(t)}$ (the overbar indicates an average over many measurement trajectories) for several values of $N$, where we observe that the QFI comes close to its maximum value $F_Q= N^2$ after $n_t\approx 200$ time steps.  This number for $n_t$ is basically independent of $N$.  In the inset of Fig.~\ref{fig2}, for $N=5$, we illustrate the convergence behavior of the QFI both for individual measurement trajectories and for the average.  

Next, we show that the target manifold \eqref{GHZ}  is reached to good accuracy.
In Fig.~\ref{fig3}(a), for $N=5$, we show a histogram of the phase $\phi=\mathrm{arg}\left(\langle0'0'0'\cdots|\Psi\rangle \langle\Psi|1'1'1'\cdots\rangle\right)$ measured after 500 time steps.  We find that the histogram is rather flat, implying that the quantum state trajectories uniformly explore the entire manifold even though they all start from the same initial state.  Alternatively, as illustrated in Fig.~\ref{fig3}(b), one may take an individual measurement trajectory and follow it over time.  For our choice of steering operators \eqref{steeringop}, almost regular oscillations are observed, where again all possible values of $\phi$ are reachable over the course of time. By invoking a termination policy, one can then target a specific state with a predesignated value of $\phi$.  
Finally, in Fig.~\ref{fig3}(c), we show the purity \cite{Schumacher_1996,Lloyd_1997,Fan2024}, 
\begin{equation}\label{purity}
    P(t)={\rm Tr}\left(\overline{\rho(t)}^2\right), 
\end{equation}
as a function of $n_t$.  Interestingly, the average state $\overline{\rho(t)}$ first approaches an
infinite-temperature state where the purity gap (almost) closes \cite{Buchhold2021}, but the purity then increases again towards the asymptotic value $1/2$. 
 This value is readily explained by the fact that averaging over the phase in Eq.~\eqref{GHZ}, one produces the asymptotic average state 
\begin{equation}\label{asmx}
    \overline{\rho}=\frac12\left(|000\cdots\rangle\langle 000\cdots|+|111\cdots\rangle\langle 111 \cdots|\right)  ,
\end{equation}
which has purity $P=1/2$ for all $N$. The difference to the numerically observed asymptotic purity in Fig.~\ref{fig3}(c) is due to the finite value of $J\delta t$. We find that this difference becomes smaller by reducing $J\delta t$, see also below.

\begin{figure}
    \centering
    \includegraphics[width=\linewidth]{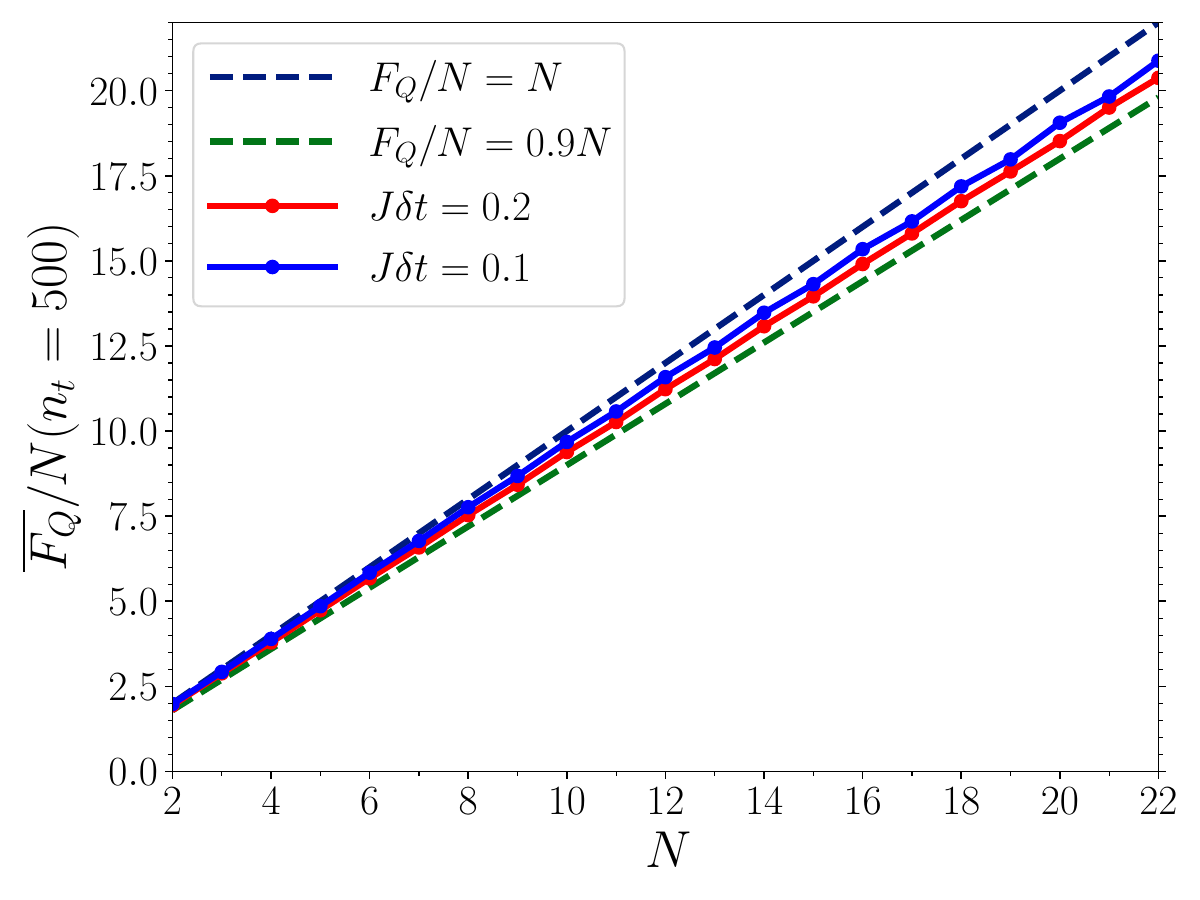}
    \caption{Scaling of the accuracy of the asymptotic value of the averaged QFI $\overline{F_Q}/N$ vs  $N$  for $J\delta t=0.1$ (blue) and $J\delta t=0.2$ (red).  Results have been averaged over at least $100$ measurement trajectories. \new{(For $N\le 10$, we used 10000 trajectories; 
    for $N=11$ to $15$, we averaged over 5000, and for $N\in\{16,17\}$, over 1000 trajectories.)}
    Solid lines are guides to the eye only. The upper bound, $\overline{F_Q}/N=N$, is shown as black dashed line, the green dashed line indicates $\overline{F_Q}/N=0.9N$.  }
    \label{fig4}
\end{figure}

\begin{figure}
    \centering
    \includegraphics[width=\linewidth]{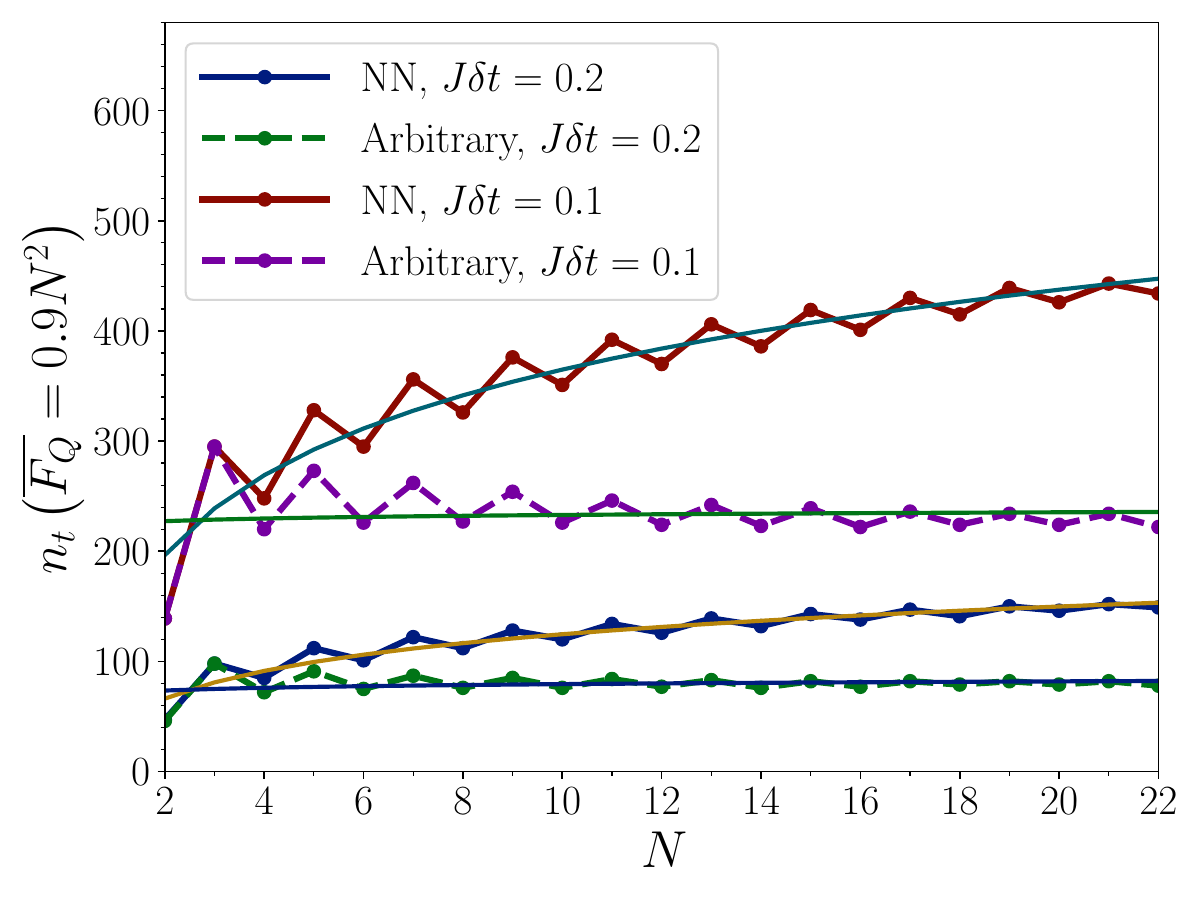}
    \caption{Scaling of the step number $n_t$ vs $N$ needed for reaching the averaged QFI value $\overline{F_Q}=0.9N^2$, shown for $J\delta t=0.1$ and $J\delta t=0.2$. Averages are over at least $100$ trajectories. \new{Thick solid and dashed curves connecting data points are guides to the eye only.
    The thin solid lines show numerical fits to the function $n_t^*(N)=A+B\ln(N)$ to the respective data, with fitting parameters $A$ and $B$.  For instance, we obtain $A\simeq 104.7$ and $B\simeq 123.9$ for nearest-neighbor couplings with $J\delta t=0.1$ from such a fit.
    We show results for nearest-neighbor detector qubit measurements (solid) as well as allowing for arbitrary detector pair measurements (dashed curves), see Fig.~\ref{fig1}.} }       
    \label{fig5}
\end{figure}

Let us now discuss the dependences on the protocol's parameters. 
The scaling of the accuracy of the value of $\overline{F_Q}$ reached after $n_t=500$ steps with system size $N$ is shown in Fig.~\ref{fig4} for two values of $J\delta t$.  We observe that for smaller $J\delta t$, higher accuracy can be reached because of the decreased importance of quantum jumps at long protocol times, which tend to deteriorate the QFI momentarily.  However, using smaller values for $J\delta t$ comes with longer physical run-times of the protocol. 

In Fig.~\ref{fig5}, we report the scaling of the number of steps $n_t$ needed for reaching a QFI of $\overline{F_Q}/N^2=0.9$, for the same values of $J\delta t$ as in Fig.~\ref{fig4}.  Here, we also compare to a situation where one allows for Bell pair measurements \new{between arbitrary detector pairs (not only nearest neighbors), see Fig.~\ref{fig1}(c), which tends to accelerate the protocol.  
In this case, all non-overlapping pairs $(n,n')$ with $n,n'\in\{1,\dots,N\}$ can again be steered simultaneously, where both the pairings (and the idle qubit for odd $N$) are chosen from a uniform random distribution for each time step.  Such a fully connected Bell measurement pairing scheme results in a faster convergence to the target state since more options for finding optimal feedback Hamiltonians can be explored.}

\new{
We observe an even-odd effect in Fig.~\ref{fig5}, in particular for small $N$, which originates from the presence of the idle qubit for odd $N$ in our Bell pair measurement scheme. Remarkably, the required number of steps $n_t$ increases only very slowly with $N$, suggesting \emph{scalability} of the active steering protocol for large $N$.  The fitting curves shown as thin solid lines in Fig.~\ref{fig5} indicate a scaling ${n_t\sim {\cal O}(\ln N)}$. 
Such favorable scaling} is also corroborated by Fig.~\ref{fig2}, where we show the $n_t$-dependence of $\overline{F_Q}$ for several (large) values of $N$.  

\begin{figure}
    \centering
    \includegraphics[width=\linewidth]{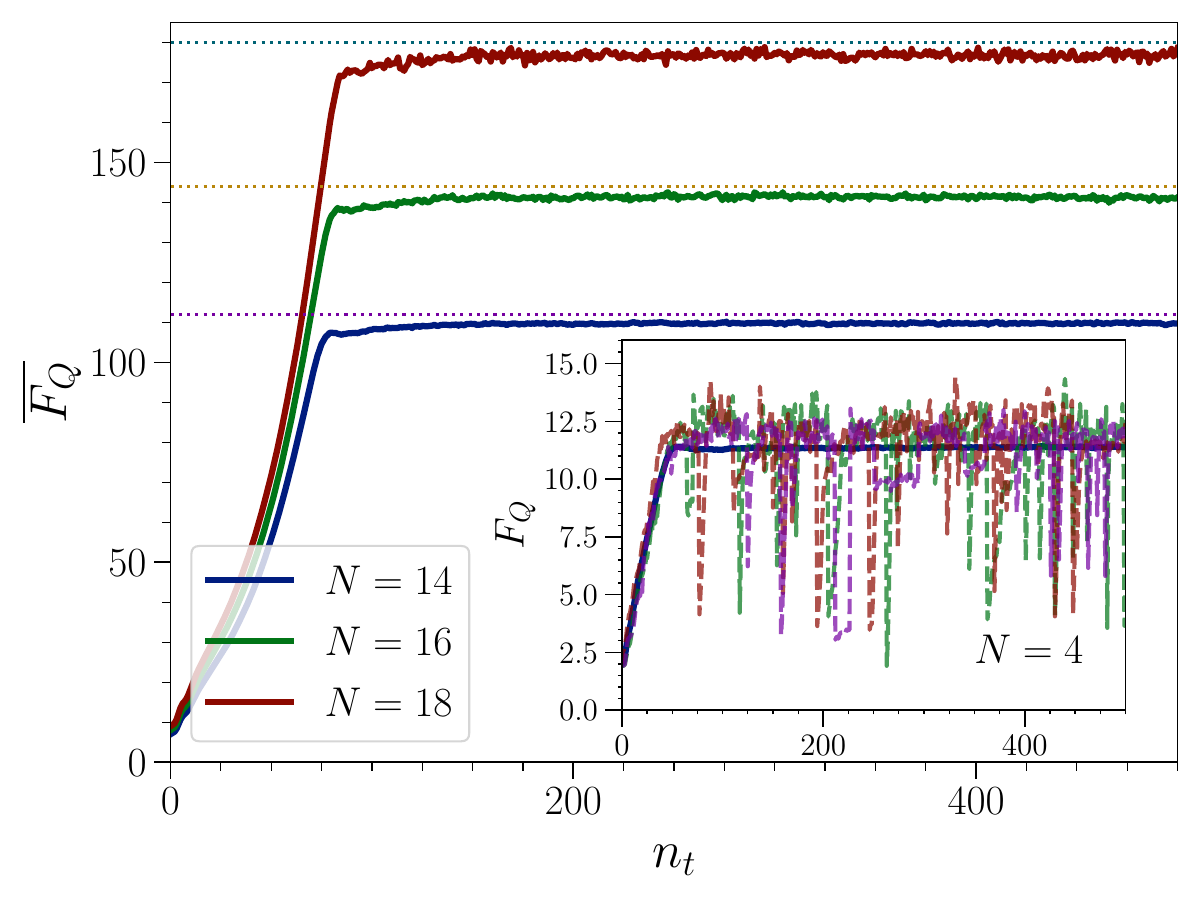} 
    \caption{\new{Average QFI $\overline{F_Q}$ vs number of time steps $n_t=t/\delta t$ for an active steering protocol with cost function $|F_Q-F_Q^\ast|$ and $F_Q^\ast$ in Eq.~\eqref{fqdicke}, targeting the Dicke states $|D_{k,N}\rangle$ in Eq.~\eqref{dicke} for $k=N/2$ and various $N$, with $J\delta t=0.2$. Averages are over $5000 \, (1000, 500)$ trajectories for $N=14 \, (16, 18)$.  Dotted lines indicate the respective $F_Q^\ast$ in Eq.~\eqref{fqdicke}. The inset shows $\overline{F_Q}$ vs $n_t$ (solid curve, averaged over $10^4$ trajectories) for $N=4$, together with three individual measurement trajectories (dashed curves).} }       
    \label{fig6}
\end{figure}

\new{In addition to targeting states with maximum QFI, our protocol also allows for the preparation of states associated with different (non-maximal) target values $F_Q^*$ of the QFI.
In order to do so, we simply choose $|F_Q-F_Q^\ast|$ as cost function in the steering protocol. As an example for this approach, 
we here consider the case of \emph{Dicke states} \cite{Pezze2018}, which for qubits can be written as
\begin{equation}\label{dicke}
    |D_{k,N}\rangle=\left( \begin{array}{c} N\\ k\end{array} \right)^{-\frac{1}{2}}\sum_{j}\mathcal{P}_j(|0\rangle^{\otimes N-k}|1\rangle^{\otimes k}).
\end{equation}
In Eq.~\eqref{dicke}, the sum runs over all possible permutations ${\cal P}_j$ of distributing $k$ excited qubits in an $N$-qubit system. 
For the state in Eq.~\eqref{dicke},  the QFI has the value  \cite{Pezze2018}
\begin{equation}\label{fqdicke}
    F_Q^\ast=\frac{N^2}{2}-2\left(\frac{N}{2}-k\right)^2+N.
\end{equation} 
As for GHZ states, the Dicke states in Eq.~\eqref{dicke} can in addition contain arbitrary phase differences between the corresponding basis states,
with the same value of $F_Q^\ast$.
In general, our protocol will therefore target an entire state manifold for a given QFI value.}

\new{In analogy to Fig.~\ref{fig2}, Figure~\ref{fig6} shows  numerical simulation results for the QFI-based preparation of Dicke states with target value
$F_Q^\ast$ in Eq.~\eqref{fqdicke} for $k=N/2$ and up to $N=18$ qubits.
We observe that the target value of the QFI is reached after $n_t\approx 70$ time steps, 
independently of the value of $N$.  The protocol thus converges significantly faster than for GHZ states. However, even though $\overline{F_Q}$ is close to the target value $F_Q^*$, individual trajectories now exhibit stronger fluctuations, see the inset of Fig.~\ref{fig6}.  We note in passing that by implementing a stoppage criterion in the protocol \cite{Morales2024}, i.e., by terminating the protocol once the QFI target value has approximately been reached, we expect that such fluctuations can be reduced.  In any case, we conclude that the QFI-based protocol is also useful for generating other highly entangled state manifolds beyond GHZ states.}

\section{Discussion}\label{sec4}

We have proposed an active steering protocol targeting the one-parameter manifold of $N$-qubit GHZ states \eqref{GHZ} with genuine multipartite entanglement by means of weak Bell pair measurements and active feedback.  In contrast to fidelity-based cost functions \cite{Morales2024}, by using the QFI as a cost function, our results suggest that the active steering protocol \new{may become} scalable with increasing system size $N$. 
Although it is well known that GHZ states may be generated by projective measurements in shallow circuits \cite{Kam2024,Sahay2025a,Sahay2025b}, our results are relevant for at least two reasons
\new{beyond those specified in Sec.~\ref{sec1}}: 
(i) We clarify
similarities and differences between active and passive steering protocols for the case of a target state \emph{manifold}.
In passive steering \cite{Roy2020,Edd2023}, measurement outcomes are discarded. Such protocols are basically equivalent to driven-dissipative systems \cite{Poyatos1996,Diehl2008,Verstraete2009,Barreiro_2011,Krauter2011}  with engineered dissipation, see Refs.~\cite{Paz1998,Barnes2000,Ahn2002,Ahn2003,Sarovar2004,Oreshkov2007,Wiseman2010,Kerckhoff2010,Kapit2016,Kapit2018,Gau2020a,Gau2020b,Lieu2020,Lieu2023,Shtanko2023,Kristensen2023} and  corresponding experiments ~\cite{Leghtas2013,Minev2019,Campagne2020,Gertler2021,Livingston2022,Lachance2024}.
In contrast to the passive steering case, where the initial state uniquely determines the final state 
within the dark space forming the target manifold \cite{Zanardi2014}, we find that under active steering,
the quantum state trajectory continues cycling through the target manifold.
The initial state then plays no special role, and all phases $\phi$ in Eq.~\eqref{GHZ} are reached with equal probability as the protocol evolves.   (ii)  One may also target highly entangled non-stabilizer state manifolds, for which simpler routes along the lines of Ref.~\cite{Kam2024} are not available. Under quite general conditions, many highly entangled states, including Eq.~\eqref{GHZ},  cannot be reached by driven-dissipative and/or passive steering  \cite{Ticozzi2012}. For small $N$, it has been demonstrated 
that they remain accessible to active steering protocols \cite{Morales2024}.

\new{In fact, one can apply a slightly modified version of the present protocol, which has favorable scaling properties with system size, in order to prepare broad classes of other states,  
including the so-called $W$, Dicke, and cluster states} \cite{Nielsen2000}. 
This task may be achieved, for instance, by stopping the protocol at a suitable value of $F_Q$ before the convergence to the maximal QFI has been reached. \new{Alternatively, one can select a target value $F_Q^\ast$ for the QFI associated with the desired family of target states, and then choose $|F_Q-F_Q^\ast|$ as cost function. For instance, the Dicke states $|D_{k,N}\rangle$ in Eq.~\eqref{dicke} with associated QFI $F_Q^\ast$ in Eq.~\eqref{fqdicke} have been studied for $k=N/2$ excitations above, see Fig.~\ref{fig6}.
Another possibility to reach non-GHZ states is to add a non-stabilizerness quantifier  \cite{Ahmadi2024,Tarabunga2024} to the cost function.}
 
\new{For experiments, a modified implementation of our protocol where state tracking is not required would be highly beneficial.  Without state tracking, one could  both study larger system sizes and tolerate error channels with weak error rates.  Such protocols can be formulated if weak measurements of all one- and two-body correlations present in the system state can be performed~\cite{Wiseman2010} since $\overline{dF_Q}$ in Eq.~\eqref{dfq} depends only on those correlations.  As a result, the protocol could be entirely based on the outcomes of weak measurements without any need for state tracking. 
In particular, our simulations only use the values of the one- and two-body correlations $R_{\mu_i}$ and $Q_{n,m}^{\alpha_n,\alpha_m}$ defined after Eq.~\eqref{eqApp:expQFIchg} in order to determine the optimal couplings $K$ entering the feedback Hamiltonian.  Counting the number of one- and two-body correlators, we expect that the performance of such a protocol will
scale $\sim N^2$.
Noting that our numerical simulations are mainly limited by the computational demands of updating the SSE, which is not necessary anymore when switching to a protocol directly utilizing weak measurements of  one- and two-body correlations, we expect that such a protocol can be applied to larger values of $N$.  In addition, it allows for the presence of error channels and/or measurement inefficiencies.  However, a detailed exploration of this interesting direction is beyond the scope of this paper.
}

\new{Finally, given that the state manifolds that can be targeted by our protocol could allow one to overcome classical limits in quantum phase estimation and to attain the so-called Heisenberg limit \cite{paris2009quantum,pezze2009entanglement,Toth2014,Liu2020}, it is a promising option to employ active steering protocols using the QFI for entanglement-enhanced metrology and sensing applications, see also Refs.~\cite{Davis2016,Macri2016,Pezze2018,Dooley2023}.  
However, this remains feasible provided errors have a limited impact and if an optimal measurement, dependent on the state, can be performed. A comprehensive analysis of how errors affect state preparation, sensing, and measurement is left for future work.
}\\

\new{
\emph{Data availability.}
The data underlying the figures and the source code used for generating the simulations are available at \cite{zenodo, github}.
}\\
 
\begin{acknowledgments} 
We thank S. Diehl, Y. Gefen, and I. Gornyi for discussions.
We acknowledge funding by the Deutsche Forschungsgemeinschaft (DFG, German Research Foundation) under Projektnummer 277101999 - TRR 183 (project B02), and under Germany's Excellence Strategy - Cluster of Excellence Matter and Light for Quantum Computing (ML4Q) EXC 2004/1 - 390534769.
\end{acknowledgments}

\bibliography{biblio}

\end{document}